\documentclass{elsart}
\usepackage{graphicx}
\usepackage{subfigure}
\usepackage{epsfig}
\usepackage{amssymb}
\begin{document}
\begin{frontmatter}
\title{Detection of Liquid Xenon Scintillation Light with a Silicon Photomultiplier}
\author[cu]{E. Aprile}, \author[umn]{P. Cushman}, \author[cu]{K. Ni},
\author[umn]{P. Shagin\corauthref{me}}
\address[cu]{Physics Department and Columbia Astrophysics Laboratory, Columbia
  University, New York, NY 10027, USA}
\address[umn]{School of Physics and Astronomy, University of Minnesota,
  Minneapolis, MN 55455, USA}
\corauth[me]{Corresponding author: tel: 1 612 625-5810; fax: 1 612 624-4478; e-mail: shagin@physics.umn.edu}
\begin{abstract}
We have studied the feasibility of a silicon photomultiplier (SiPM) to detect liquid xenon 
(LXe) scintillation light. The SiPM was operated inside a small volume of pure LXe, 
at -$95\rm{^oC}$, irradiated with an internal $^{241}$Am $\alpha$ source. The gain of 
the SiPM at this temperature was
estimated to be $1.8 \times 10^6$ with bias voltage at 52 V. Based on the geometry of the setup, 
the quantum efficiency of the SiPM was estimated to be 22$\%$ at the Xe wavelength of 178 nm. 
The low excess noise factor, high single photoelectron
detection efficiency, and low bias voltage of SiPMs make them attractive
alternative UV photon detection devices to photomultiplier tubes (PMTs) for liquid 
xenon detectors, especially for experiments requiring a very low energy detection threshold, 
such as neutralino dark matter searches. 
\end{abstract}
\begin{keyword}
SiPM \sep Liquid Xenon \sep Dark Matter
\PACS 
\end{keyword}
\end{frontmatter}
\section{Introduction}
\label{}

The SiPM\cite{sipm} is a promising Avalanche Photodiode (APD) variant consisting of 
576 silicon micro pixels per square mm of detector surface. Each pixel is 
a $21 \times 21 \rm\mu m^2$ independent photon 
micro-counter operating in limited Geiger mode with gain of $10^6$. All SiPM pixels are connected 
to the common load, so  the output signal is the sum of all signals. Thus a proportional signal 
is created by the sum of the digital micro-APD (pixel) signals. The main features of SiPM are 
low excess noise 
factor, low bias voltage (50V), and excellent timing (30 ps for 10 photoelectrons).
SiPMs have a low excess noise factor comparable to the Hybrid Photodiode (HPD)\cite{hpd}
because
the gain mechanism relies on counting how many of the micro-APDs have fired. 
The SiPM noise is high at room temperatures\cite{sipm}, but is reduced significantly 
when operated at cryogenic
temperatures. The photon detection efficiency is similar to a PMT, but comes from the product of
a higher quantum efficiency (QE) multiplied by the ratio of sensitive area to the total detector 
area. It is thus well-suited to a purely solid state solution to LXe scintillation 
detection.

Liquid xenon is a very good scintillation material for various applications of particle 
detection\cite{XMASS,MEG,EXO,PET}. The Columbia group, in particular, is interested in 
development of a liquid xenon time projection chamber (TPC) for Compton imaging of MeV 
sources in high energy astrophysics \cite{LXeGRIT}, and more recently, in the development 
of a dual phase (liquid/gas) TPC for the direct detection of dark matter particles in the 
form of WIMPs (Weakly interacting massive particles)\cite{XENON}. In all these applications, 
efficient detection of the LXe scintillation is key to the sensitivity of the detector 
and its minimum energy threshold.

The wavelength of liquid xenon scintillation light is centered around 
178 nm \cite{Jortner:65}, making its detection a challenge. Currently, the TPC for the 
XENON Dark Matter Search\cite{XENON} uses compact metal channel PMTs to detect both primary 
and proportional scintillation light produced by WIMP nuclear recoils. These PMTs 
(R9288 and R8520), produced by Hamamatsu Photonics Co., have typical quantum efficiencies 
of 20\% at liquid xenon temperature ( $-95\rm^{o}C$).

In the development phase of the XENON TPC, we have tested alternative UV photon sensors in liquid 
xenon, including multi-channel plate PMTs (MCP-PMT) and large area avalanche photodiodes (LAAPDs). 
The SiPM is another very promising device, which can detect a small amount of light with 
a very good single photoelectron detection capability. Here we summarize our first attempt 
to detect liquid xenon scintillation light with a small SiPM immersed in liquid xenon.

\section{Experimental Apparatus}
\label{DAQ}

\begin{figure}[p] 
\begin{center}
\hspace{2.5pt}  
\includegraphics[width=0.8\textwidth]{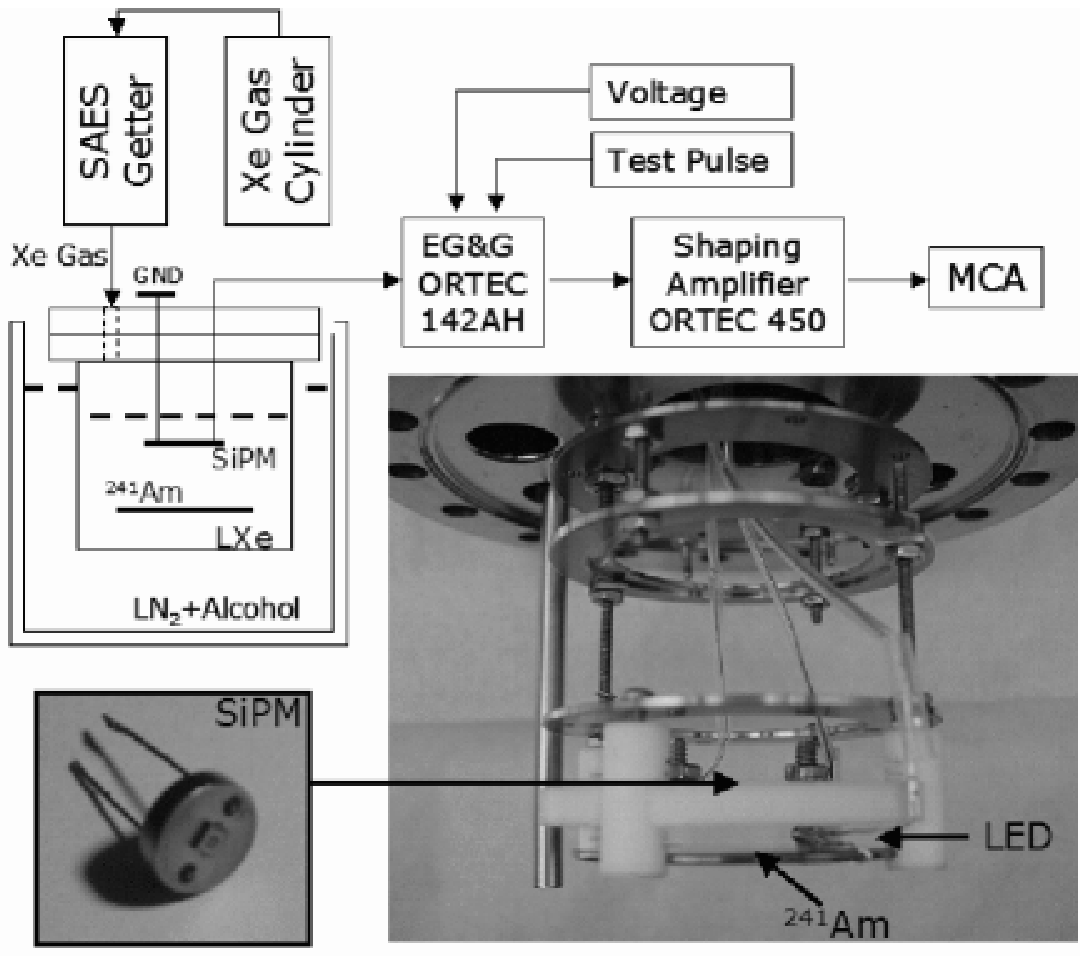}
\caption{ The chamber insert used for the Columbia Nevis test is shown above with the SiPM, 
source and LED clearly illustrated.  The whole arrangement is immersed in purified 
liquid xenon.  Included is a diagram showing the complete detector, with the gas 
filling system, and data acquisition system (DAQ).}
\label{fig:Nevis}
\end{center}
\end{figure}

\begin{figure}[p] 
\begin{center}
\hspace{2.5pt} 
\includegraphics[width=0.75\textwidth]{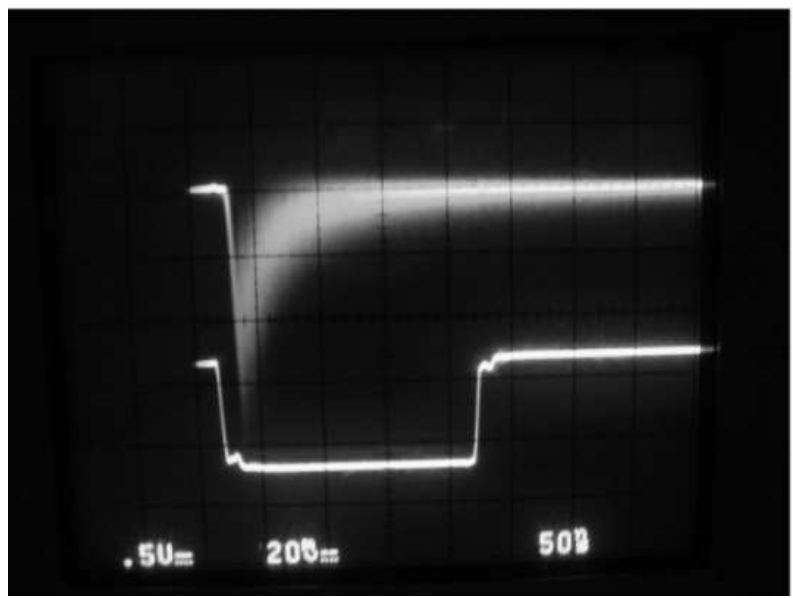}
\caption{LED signal detected by silicon photomultiplier as seen on the 
screen of oscilloscope.}
\label{fig:Signal}
\end{center}
\end{figure}

The LXe detector used for the test of the SiPM is the same one used for testing different 
light sensors, including the Hamamatsu metal channel PMT\cite{Aprile:04} and the LAAPD\cite{apd}. 
The detector consists of a 6 cm diameter stainless steel electrode with a radioactive alpha 
source $^{241}$Am deposited in its center. The $1 \times 1\rm mm^2$ SiPM (type Z, serial 
number 217) was mounted on a Teflon support plate facing the source plate. A blue LED was 
also mounted on the Teflon plate. The distance between source and SiPM was 4.7 mm. The 
detector was pumped down to a vacuum level of \(10^{-8}~\rm{Torr}\) and baked out for 36 
hours before filling with LXe. Xe gas, purified through a SAES 
getter\footnote{http://www.saesgetters.com/}, was condensed in the detector vessel, cooled 
by a bath of liquid nitrogen and alcohol mixture at $-95\rm{^oC}$. Fig.\ref{fig:Nevis} 
shows a schematic of detector, gas system and electronics. A scope trace of the blue LED 
signal as detected by SiPM is shown in Fig. \ref{fig:Signal}.

The scintillation photons, absorbed by the SiPM, produce photoelectrons which are consequently amplified 
inside the silicon via a Geiger mode avalanche. The SiPM electrical signal is fed into a charge sensitive pre-amplifier, followed by an ORTEC 450 shaping amplifier. A test pulse generator is used to calibrate electronics chain system. 

\section{Results}

\subsection{Calibration}

\begin{figure}[ht] 
\begin{center}
\hspace{2.5pt} 
\includegraphics[width=0.9\textwidth]{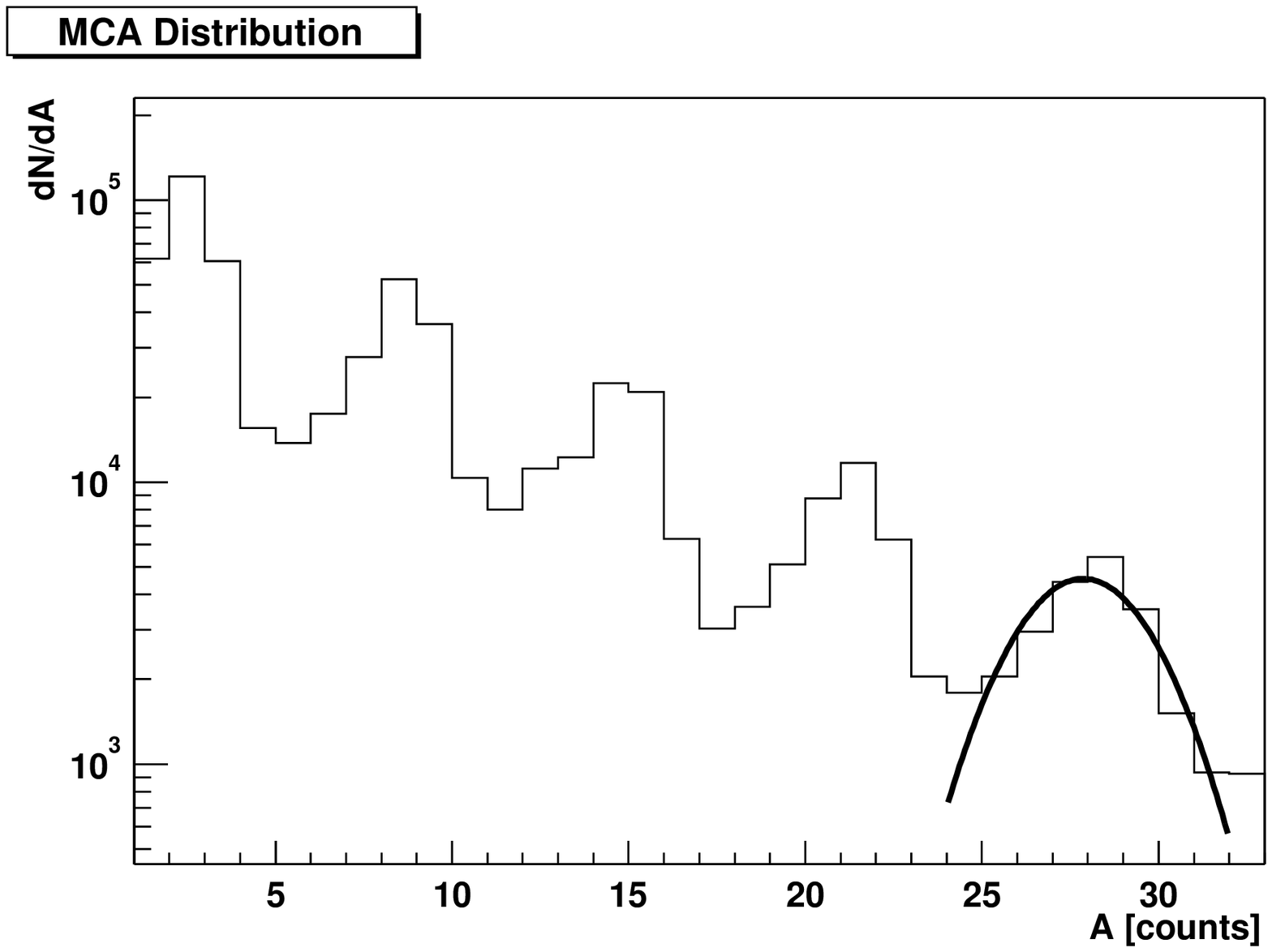}
\caption{Amplitude distribution for $^{241}$Am particle scintillations. 
Low amplitude part of the spectra.}
\label{fig:LowAmp}
\end{center}
\end{figure}

\begin{figure}[h] 
\begin{center}
\hspace{2.5pt} 
\includegraphics[width=0.9\textwidth]{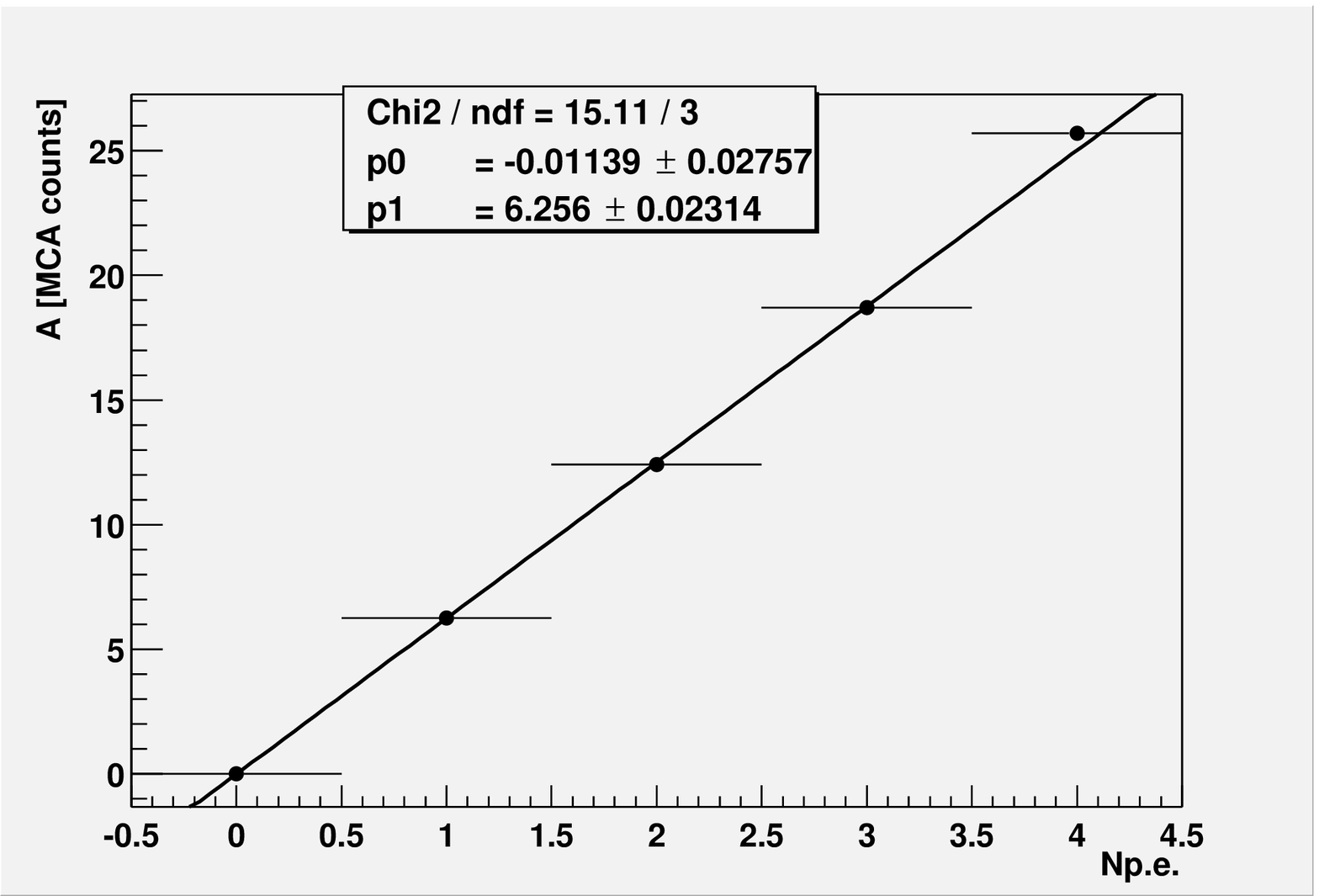}
\caption{Mean amplitude after pedestal subtraction versus peak number. Fit - linear function. The MCA offset from the channel number of zero photoelectron peaks in Fig.\ref{fig:LowAmp} is subtracted.}
\label{fig:Calibr}
\end{center}
\end{figure}

The great advantage of a SiPM is that it is self-calibrating, since its single photoelectron sensitivity can be used for calibration. The resulting low amplitude part of the $\alpha$-source spectrum is shown in Fig. \ref{fig:LowAmp}. Note the excellent resolution 
(low excess noise factor) that allows up to 11 photoelecton peaks to be clearly 
distinguished. Each single photoelectron peak is fitted using a Gaussian function. The fitted mean value is plotted versus peak number (number of photoelectrons) in 
Fig. \ref{fig:Calibr}, clearly showing a linear behavior.


\subsection{Gain measurement}
We have estimated the SiPM gain $g$ with a calibrated test pulse signal as follows:
\begin{center}
\begin{equation}
g = \frac{V_{t} C}{q A} \cdot \frac{M_1}{M_t}
\end{equation}
\end{center}

where $V_{t}$ and $M_{t}$ are the test pulse amplitude and channel number in the MCA spectrum, respectively. $C$ is the capacitance of the charge sensitive
pre-amplifier. 
$M_{1} \approx 6.2$ is number of channels corresponding to a single photoelectron and 
$q$ is the elementary electronic charge. $A \approx 27$ is the gain of the amplification system used in 
this measurement. Based on these values, the SiPM gain is 
estimated to be around $1.8 \times 10^6$ at the operation voltage of 52 V.

\subsection{Quantum Efficiency Estimation}
\label{sec:Al}
\begin{figure}[h] 
\begin{center}
\hspace{2.5pt} 
\includegraphics[width=0.9\textwidth]{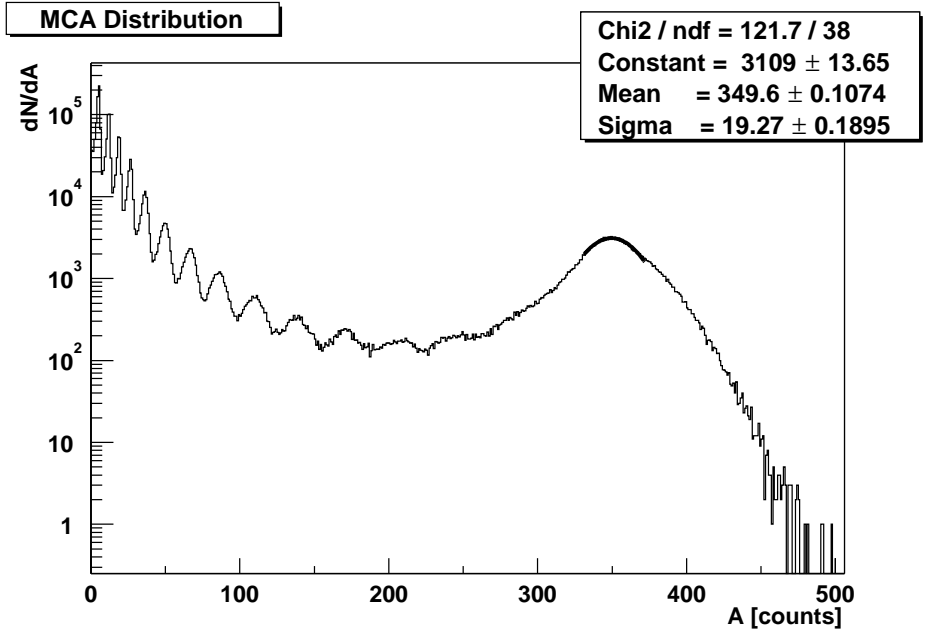}
\caption{Amplitude distribution for $^{241}$Am particle scintillations. Fit - Gaussian function.}
\label{fig:Alpha}
\end{center}
\end{figure}

In Fig. \ref{fig:Alpha} the $^{241}$Am $\alpha$-source energy spectrum, measured with the SiPM bias voltage at 52 V, is shown.

Since the SiPM signal is calibrated in units of photoelectrons, the average number of photoelectrons
detected from LXe scintillation light induced by $^{241}$Am $\alpha$ particles is determined to be 55 p.e.

The total number of scintillation photons striking the SiPM can be calculated using the energy 
of $\alpha$ particle (5.48 MeV), the average energy needed to produce single 
scintillation photon in LXe
(19.6 eV for $\alpha$ particle \cite{Doke:02}), and the geometrical acceptance of the SiPM detector.
For the detector geometry used in the tests, the average number of scintillation photons 
produced by one 5.48 MeV $\alpha$ particle striking the SiPM surface is $N_{ph}$ = 1006 photons. 
Thus the measured photon detection efficiency  is $\varepsilon = \frac{55 p.e.}{1006 ph.} =  5.5\%$. 
The quantum efficiency of the SiPM (QE) can be calculated as $ \varepsilon  = QE \times A$, where $A$ 
is the active area 
ratio of the device. Assuming $A$ = 0.254\cite{sipm}, we infer a $QE$ = 22$\%$ including the probability of initiating the Geiger avalanche.

\section{Conclusion}

A silicon photomultiplier was tested for the first time in LXe to detect its scintillation 
light at $\lambda$ = 178 nm, at an operation temperature at $-95\rm{^oC}$. A high quantum 
efficiency of 22\% has been demonstrated. Large arrays of SiPMs offer a promising solid state 
photodetector approach for reading out LXe detectors in applications ranging from $\gamma$-ray 
astrophysics to particle physics and medical imaging.

\section{Acknowledgments}

The authors wish to express their gratitude to Prof. B.Dolgoshein and Dr.
E.Popova for the donation of SiPM samples and their assistance in understanding 
SiPM behavior. This work was supported in part by grants from the National Science 
Foundation to the Columbia Astrophysics Laboratory (Grant No. PHY-02-01740) and 
by The Department of Energy (HEP grant No. 94ER40823) to the University of Minnesota.

\end{document}